\documentclass{article}
\usepackage{spconf,amsmath,graphicx,hyperref}
\usepackage{float}      
\usepackage{multirow}   
\usepackage{booktabs}   
\usepackage[table]{xcolor} 
\usepackage{algorithm}
\usepackage{algorithmic}
\usepackage{booktabs} 
\usepackage[utf8]{inputenc} 
\usepackage{fancyhdr}

\title{SimulatorCoder: DNN Accelerator Simulator Code Generation and Optimization via Large Language Models}
\name{Yuhuan Xia, Tun Li\textsuperscript{*}, Hongji Zhou, Xianfa Zhou, Chong Chen, Ruiyu Zhang\thanks{*Corresponding author.}\vspace{-2.5ex}}

\address{College of Computer Science and Technology, National University of Defense Technology, Changsha, China \\
Email: \{xiayuhuan, tunli, zhouhongji, zhouxf23, chenchong, zhangruiyu\}@nudt.edu.cn}

\begin{document}
%
\fancypagestyle{firstpage}{
  \fancyhf{}
  \renewcommand{\headrulewidth}{0pt}   
  \fancyhead[C]{\small To appear on ICASSP'26. This paper is a draft.}
}

\thispagestyle{firstpage}

\maketitle
\begin{abstract}
This paper presents SimulatorCoder, an agent powered by large language models (LLMs), designed to generate and optimize deep neural network (DNN) accelerator simulators based on natural language descriptions. By integrating domain-specific prompt engineering including In-Context Learning (ICL), Chain-of-Thought (CoT) reasoning, and a multi-round feedback-verification flow, SimulatorCoder systematically transforms high-level functional requirements into efficient, executable, and architecture-aligned simulator code. Experiments based on the customized SCALE-Sim benchmark demonstrate that structured prompting and feedback mechanisms substantially improve both code generation accuracy and simulator performance. The resulting simulators not only maintain cycle-level fidelity with less than 1\% error compared to manually implemented counterparts, but also consistently achieve lower simulation runtimes, highlighting the effectiveness of LLM-based methods in accelerating simulator development. Our code is available at \url{https://github.com/xiayuhuan/SimulatorCoder}.
\end{abstract}
\section{Introduction}
\label{sec:intro}

With the rapid development of artificial intelligence, deep neural network (DNN) accelerators have been widely used in important and complex scenarios such as cloud computing and edge inference \cite{9534784}. Due to the increasing complexity of DNN accelerator design, researchers are seeking tools that enable rapid and accurate design space exploration during the early stages of design and to quickly evaluate the effects of architectural improvements \cite{9651675}. Currently, evaluation methods are primarily based on analytical models \cite{9076333} or simulators \cite{samajdar2020systematic, 9668279}. Analytical models use mathematical formulas and simplifications which often fail to capture subtle execution differences, resulting in a low accuracy of evaluation results. As DNN accelerator architectures become more complex and flexible, the accuracy and reliability of analytical models significantly degrade. DNN accelerator simulators aim to faithfully reproduce the cycle-level execution behavior of neural networks when mapped onto hardware accelerators. They capture how computations are partitioned, scheduled, and executed across processing elements. However, building a dedicated simulator for each new DNN accelerator to be designed is time-consuming and resource-intensive.

The emergence of large language models (LLMs) has brought new opportunities for automatic chip design. LLMs can be used to generate code based on natural language descriptions, and many approaches have been proposed to improve the accuracy of LLM-based code generation, including prompt engineering, fine-tuning \cite{10.1145/3714462}, and program self-repair \cite{10.1145/3691620.3695506}. Prompt engineering techniques include In-Context Learning (ICL) \cite{10.1145/3715908}, Chain-of-Thought (CoT) \cite{10.1145/3690635} and related strategies. Based on these technologies, existing studies \cite{10323953, vungarala2025tpugenllmdrivencustomtensor} have begun to explore the use of natural language to drive AI chip design, achieving promising initial results.  GPT4aigChip \cite{10323953} proposes an LLM-friendly hardware template and a demonstration-enhanced prompt generator for the automatic design of AI accelerators.


However, LLM-based DNN accelerator simulator code generation and optimization still face several key challenges, including the lack of domain knowledge prompts and high complexity of implementation. Existing prompt engineering methods often lack sufficient domain-specific guidance. Without structural or template-based prompting, the generated code frequently lacks proper alignment with hardware semantics. Simulators aim to faithfully model the computing behavior of accelerators. However, unlike the mature and well-established design toolchains available for DNN accelerators, simulator design is usually a self-built framework, and the verification process still frequently depends on extensive manual work, which limits scalability and efficiency and ultimately poses even greater challenges.

\begin{figure}[htb]

\begin{minipage}[b]{1.0\linewidth}
  \centering
  \centerline{\includegraphics[width=8.5cm]{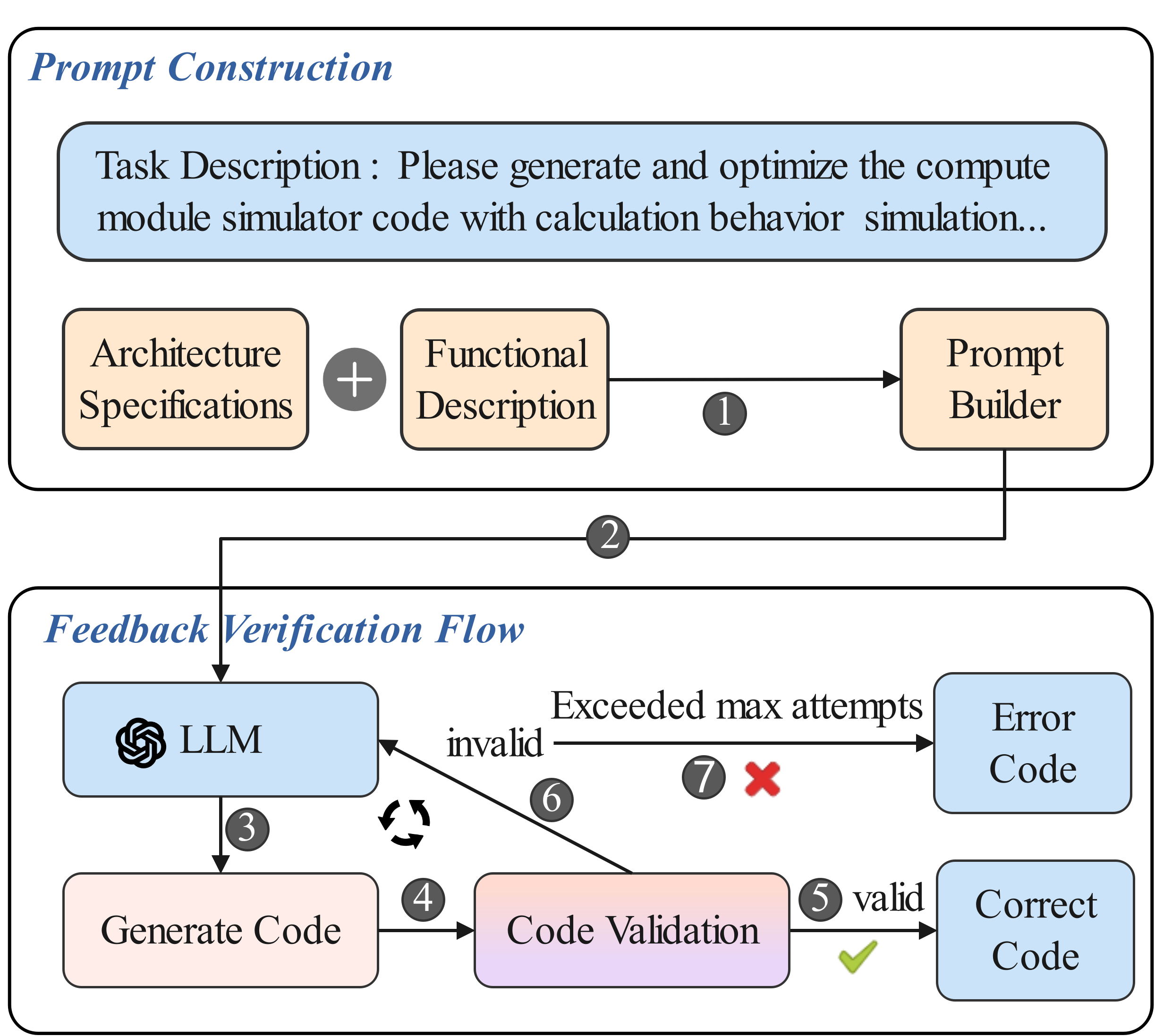}}
\end{minipage}
\caption{Overview of the SimulatorCoder framework. The framework consists of two core modules: prompt engineering, which guides the LLM with systematic and domain-specific prompts, and feedback verification, which iteratively refines and validates generated code to ensure functional correctness and efficiency.}
\label{fig:framework}
\end{figure}

To address the limitations of manually designed traditional simulators, we introduce SimulatorCoder, a new untrained large language model agent capable of generating and optimizing simulator code based on functional descriptions. Our main contributions are as follows: (1) To the best of our knowledge, this work is the first to generate and optimize DNN accelerator simulator code using LLMs and to develop an agent framework, SimulatorCoder. (2) We summarize complex design requirements and design context-rich prompt templates leveraging CoT and ICL techniques. These elements guide SimulatorCoder in generating semantically correct and structurally complete modules. (3) We establish a feedback verification flow. SimulatorCoder leverages error feedback to iteratively refine and optimize the generated simulator code. Experiments on the customized SCALE-Sim \cite{samajdar2020systematic} benchmark show that SimulatorCoder achieves high code generation accuracy and is capable of simulator optimization.


\section{METHODOLOGY}
SimulatorCoder is an LLM-based agent that accurately converts functional requirements described in natural language into executable simulator code. Figure \ref{fig:framework} shows the core framework of SimulatorCoder, which consists of two key modules: prompt engineering and feedback verification flow. (1) The prompt engineering module guides the LLM to conduct systematic design thinking and improve the logical structure of code generation by designing a series of strategic problem-solving prompts. (2) The feedback verification module adopts an inspection mechanism to detect and correct errors in the generated code. It provides detailed diagnostic feedback, enabling the LLM to iteratively optimize the generated code design to ensure the functionality and accuracy of the final output code. Our approach is outlined in Algorithm \ref{alg:algorithm}, with further details provided in the following sections.

\subsection{Simulator Template}
We decompose the core functionality of the simulator into three modules: the mapping module, the storage module, and the interconnection network module. The mapping module is responsible for mapping the layer operations of the neural network to the computing unit. The storage module includes the simulation of on-chip and off-chip storage, and is responsible for managing data storage and access. The interconnection network module simulates data transmission within and between the computing unit and the external storage. To construct the simulator, we adopt a modular decomposition and incremental construction strategy, generate code by module, and incrementally build function-level, class-level, and module-level code generation. We use prompt engineering to enable LLM to learn domain knowledge.

\begin{algorithm}[tb]
\caption{SimulatorCoder Agent Framework}
\label{alg:algorithm}
\textbf{Input}: Task $T$, Architecture Specifications $AS$.\\
\textbf{Output}: \makebox[0pt][l]{Simulator design code $C_K$, Simulation results $SR$.}
\begin{algorithmic}[1] 
\STATE Initialize LLM using control instructions.
\WHILE{design code task not complete}
\STATE Given $AS$ and task $t_j$, form $promp_j$= $AS + t_j$.
\STATE Generate simulator code $C_K$.
\STATE Generate simulation results $SR$.
\WHILE{in max $attempt$}
\STATE Evaluate $C_K$ via compilation and simulation.
\IF {$C_K$ is valid}
\STATE return $\{C_K, SR\}$
\ELSE
\STATE $C_K \leftarrow New C_K$.
\STATE $SR \leftarrow New SR$.
\ENDIF
\ENDWHILE
\ENDWHILE
\STATE \textbf{return} $\{C_K, SR\}$
\end{algorithmic}
\end{algorithm}

\begin{figure}[htb]

\begin{minipage}[b]{1.0\linewidth}
  \centering
  \centerline{\includegraphics[width=8.5cm]{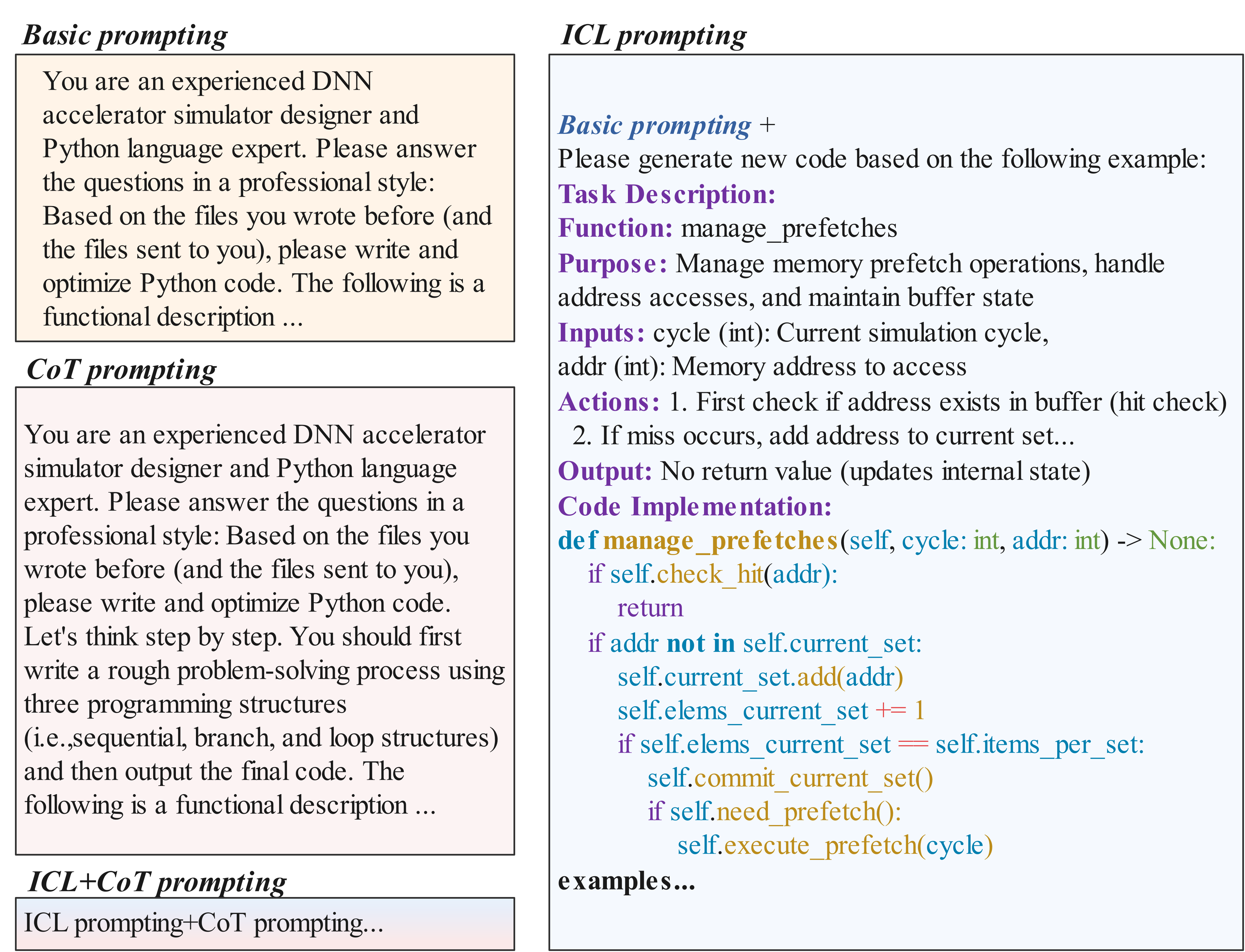}}
\end{minipage}
\caption{Examples of domain-specific prompting approaches in SimulatorCoder, including basic (Zero-shot), ICL, CoT, and the integrated ICL and CoT strategy.}
\label{fig:prompt}
\end{figure}

\subsection{Prompt Engineering}
For complex simulator design tasks, LLMs are often required to capture domain knowledge. Our approach is based on ICL and CoT reasoning techniques. ICL adds input-output examples to the prompt, enabling LLMs to solve new problems through few-shot learning by imitating existing examples. CoT reasoning requires the model to explicitly generate intermediate step-by-step reasoning before outputting the final answer, thereby reducing complexity. DNN accelerator architecture specifications and simulator function descriptions are manually summarized. As shown in Figure \ref{fig:framework}, the prompt builder standardizes the user's input format, conveying the intended functionality, design expectations, and architecture specifications of the simulator (step 1), and constructing prompts based on ICL and CoT techniques. Subsequently, the structured prompt is fed into the LLM (step 2).
\subsection{Feedback Verification Flow}
LLMs often generate code containing a variety of errors. Therefore, guiding LLMs to correct these issues based on error feedback is essential. As illustrated in steps 3 through 7 of Figure \ref{fig:framework}, the feedback verification flow proceeds as follows:

\textbf{LLM code generation:} The LLM receives the structured prompt and subsequently generates code based on the functional description and architectural specifications (step 3).

\textbf{Code evaluation:}  The generated code undergoes validation (step 4), including syntax and functional correctness tests. Syntax correctness is first checked via compilation. The simulator code is then executed, and its correctness is judged based on whether the output is valid. 

\textbf{Feedback self-repair:} If the evaluation result is valid (step 5), the generated code is accepted as the output. Otherwise, if the evaluation result is invalid (step 6), the LLM refines the code using the feedback. If the number of LLM code generation attempts exceeds the maximum allowed (step 7), the generation is terminated. Finally, the simulator is executed on tasks to confirm functional fidelity and performance alignment with the intended architectural specifications.

\section{Experiments}

\subsection{Experimental Design}
We investigate the capability of SimulatorCoder to generate and optimize DNN accelerator simulator code by formulating the following research questions (RQs).

\begin{table}[H]
\centering
  \caption{The Pass@k (\%) of Prompting Approaches.}
  \vspace{8pt}
  \label{tab:pass}
  \begin{tabular}{cccl}
    \toprule
    \bfseries LLMs & \bfseries Prompting & \bfseries pass@1 & \bfseries pass@5 \\
    \midrule
    \multicolumn{1}{c}{\multirow{4}{*}{DeepSeek-V3}} & Zero-shot prompting & \cellcolor{blue!5} 81.00\% & \cellcolor{blue!5} 88.00\% \\
    & ICL prompting & \cellcolor{blue!15} 88.41\% & \cellcolor{blue!15} 90.58\% \\
    & CoT prompting & \cellcolor{blue!15} 87.68\% & \cellcolor{blue!15} 92.75\% \\
    & ICL+CoT prompting & \cellcolor{blue!25} 91.30\% & \cellcolor{blue!25} 96.38\% \\
    \midrule
    \multicolumn{1}{c}{\multirow{4}{*}{GPT-4o}} & 
    Zero-shot prompting & \cellcolor{blue!5} 82.60\% & \cellcolor{blue!5} 87.68\% \\
    & ICL prompting & \cellcolor{blue!15} 89.13\% & \cellcolor{blue!15} 89.85\% \\
    & CoT prompting & \cellcolor{blue!15} 85.50\% & \cellcolor{blue!15} 89.13\% \\
    & ICL+CoT prompting & \cellcolor{blue!25} 91.30\% & \cellcolor{blue!25} 92.75\% \\
    \bottomrule
  \end{tabular}
\end{table}

\textit{RQ1}: To what extent can SimulatorCoder generate correct and executable code for DNN accelerator simulator?

\textit{RQ2}: How do different prompt strategies influence the quality and accuracy of the generated simulator code?

\textit{RQ3}: How does the performance of the simulator generated by SimulatorCoder compare to a manual implementation across various workloads?

Referencing previous LLM-based code generation research \cite{10.1109/ICSE48619.2023.00179}, we use self-defined Pass@$k$ as our evaluation metric. 
We calculate Pass@$k$ = $c / n$. Here, $n$ denotes the number of tasks. In each experiment, given a code generation and optimization task, the simulator is allowed to generate up to $k$ programs. If any generated program passes all test cases, the experiment is considered successful, where $c$ is the number of successful trials.


 

 


\subsection{Experimental Setup}
We initially evaluate SimulatorCoder's code generation capability using customized SCALE-Sim \cite{samajdar2020systematic} benchmark. SCALE-Sim is a cycle-accurate DNN accelerator simulator with configurable systolic arrays. Based on SCALE-Sim, we extracted the DNN accelerator architecture specifications and designed 138 tasks. We tasked the LLMs with generating Python code, leveraging their demonstrated effectiveness in Python programming \cite{10.1145/3643674}. We selected two popular LLMs: DeepSeek-V3 \cite{liu2024deepseek} and GPT-4o \cite{hurst2024gpt} to generate functionally equivalent simulator code. Figure \ref{fig:prompt} illustrates representative structured prompts produced by our prompt builder.

\begin{figure*}[htbp]
    \begin{minipage}{1.0\textwidth}
        \centering
        \includegraphics[width=\textwidth]{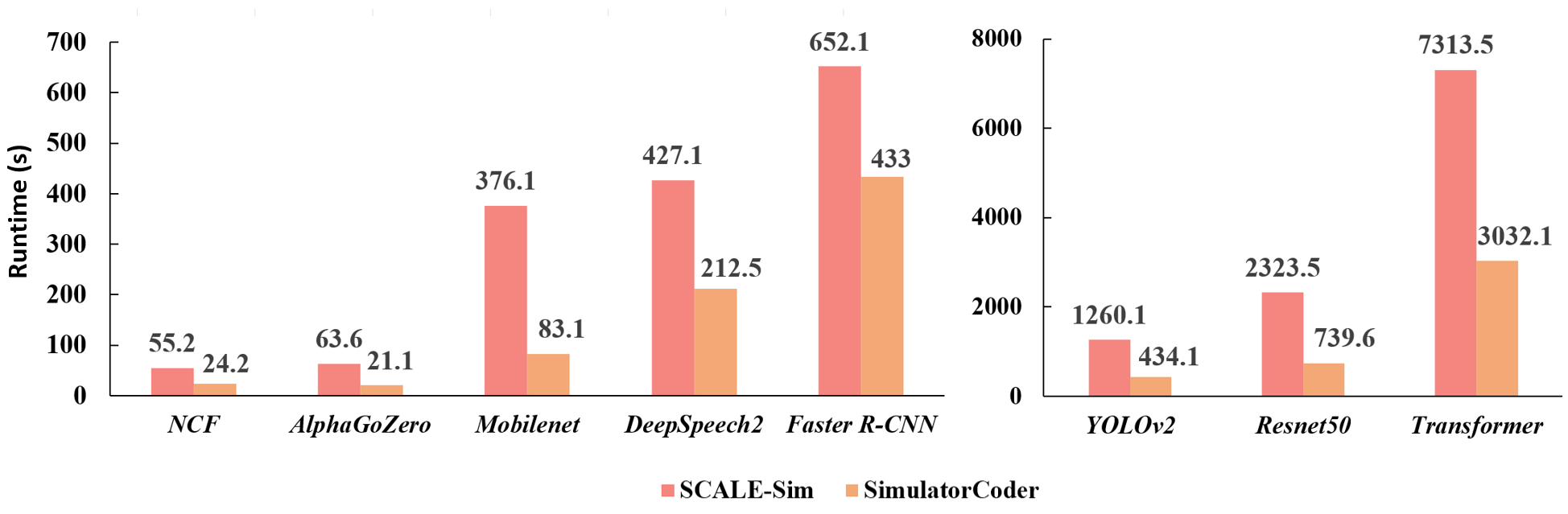}
        \setlength{\abovecaptionskip}{-0.4cm}  
        \caption{Runtime comparison between SCALE-Sim and SimulatorCoder.}
        \label{fig:Runtime}
    \end{minipage}
\end{figure*}



\subsection{Results and Analyses}
Table \ref{tab:pass} reports the Pass@$k$ results across different prompting strategies. Basic prompting strategies exhibit lower code generation accuracy, and error feedback is required to generate the correct code. Common error types include syntax errors, variable type mismatches, and function call errors. Feedback verification substantially improve the accuracy of code generation.
Importantly, SimulatorCoder is able to generate simulator code that replicates the core functionality of SCALE-Sim. Therefore, for \textit{RQ1}, we conclude that SimulatorCoder not only has the ability to generate accelerator simulator function code, but also largely ensures that the generated code is correct, executable, and closely aligned with the intended architectural specifications.

Table \ref{tab:pass} demonstrates consistent accuracy trends across both LLMs, with Pass@$k$ values showing progressive improvement from Zero-shot prompting to ICL, CoT, and finally the integrated ICL and CoT strategy. For \textit{RQ2}, we conclude that both ICL and CoT prompting strategies significantly enhance SimulatorCoder's code generation accuracy, indicating that domain-specific prompt engineering not only improves correctness but also reduces the need for extensive feedback iterations, thereby effectively accelerating the overall simulator development process.

Figure \ref{fig:Runtime} shows the runtime comparison between SCALE-Sim and SimulatorCoder across eight networks: Neural Collaborative Filtering (NCF) \cite{he2017neural}, AlphaGo Zero \cite{silver2017mastering}, MobileNet \cite{howard2017mobilenets}, DeepSpeech2 \cite{amodei2016deep}, Faster R-CNN \cite{ren2015faster}, YOLOv2 \cite{8100173}, ResNet50\cite{he2016deep}, and Transformer \cite{vaswani2017attention}. For most workloads, simulators generated by SimulatorCoder achieve shorter execution times, primarily because prompting LLMs with optimization rules yields more streamlined code that eliminates redundant operations, improves control-flow efficiency, and reduces unnecessary computational overhead, thereby accelerating simulation without sacrificing accuracy.

\begin{table}[htbp] 
    \centering
    \renewcommand{\arraystretch}{1.08} 
    \setlength{\tabcolsep}{4pt} 
    \caption{Comparison of SimulatorCoder and SCALE-Sim simulation cycles with error rate.}
    \vspace{8pt}
    \label{tab:model_cycles}
    \begin{tabular}{lcccc}
        \toprule
        \textbf{Model} & \textbf{SimulatorCoder} & \textbf{SCALE-Sim} & \textbf{Error\%} \\
        \midrule
        NCF & 552,624 & 552,616 & 0.00\% \\
        AlphaGoZero & 416,502 & 416,494 & 0.00\% \\
        MobileNet & 1,333,481 & 1,338,519 & 0.38\% \\
        DeepSpeech2 & 2,198,754 & 2,200,287 & 0.07\% \\
        Faster R-CNN & 4,329,971 & 4,367,574 & 0.86\% \\
        YOLOv2 & 2,556,983 & 2,556,974 & 0.00\% \\
        ResNet50 & 4,396,573 & 4,434,168 & 0.85\% \\
        Transformer & 4,871,474 & 4,870,117 & 0.03\% \\
        \bottomrule
    \end{tabular}
\end{table}

Table \ref{tab:model_cycles} reports the cycle count error of the simulator code generated by LLMs in comparison with the SCALE-Sim tool. The results demonstrate that, across eight representative models such as ResNet50, Faster R-CNN, and Transformer, the simulation error rates remain below 1\%. Remarkably, the error is reduced to 0.00\% for NCF, AlphaGo Zero, and YOLOv2, indicating accuracy fully consistent with SCALE-Sim. Together with the runtime comparison shown in Figure \ref{fig:Runtime}, these results highlight that LLM-generated code not only preserves accuracy but also achieves higher simulation efficiency. For \textit{RQ3}, the performance evaluation indicates that the simulators generated by SimulatorCoder achieve accuracy comparable to the manually developed simulator across diverse workloads while simultaneously reducing runtime.
\section{Conclusion}

We propose SimulatorCoder, an LLM-based agent for DNN accelerator simulator code generation and optimization. By combining domain-specific prompt engineering with ICL, CoT, and a feedback-verification flow, SimulatorCoder translates functional requirements into executable, efficient, and architecture-aligned simulator modules, including mapping, storage, and network components. Experimental results show that the generated simulators achieve high code generation accuracy, produce results comparable to the manually developed SCALE-Sim simulator across diverse workloads, and typically deliver shorter runtimes. This demonstrates that LLM-produced code can preserve cycle-level fidelity and improve overall simulation efficiency. 

\vfill\pagebreak




\clearpage
\bibliographystyle{IEEEbib}
\bibliography{strings,refs}

\end{document}